\documentclass[conference]{IEEEtran}
\IEEEoverridecommandlockouts
\usepackage{cite}
\usepackage{amsmath,amssymb,amsfonts}
\usepackage{algorithmic}
\usepackage{graphicx}
\usepackage{tabularx}
\usepackage{textcomp}
\usepackage{xcolor}
\usepackage{subcaption}
\usepackage{makecell}
\usepackage{caption}
\usepackage{bm}
\usepackage{array}
\usepackage{amsmath}
\usepackage{calc}
\usepackage{bm}
\usepackage{lipsum}
\usepackage{array,multirow,graphicx}
\usepackage[bookmarks=false]{hyperref}
\newcolumntype{P}[1]{>{\centering\arraybackslash}p{#1}}
\newcolumntype{M}[1]{>{\centering\arraybackslash}m{#1}}

\def\BibTeX{{\rm B\kern-.05em{\sc i\kern-.025em b}\kern-.08em
    T\kern-.1667em\lower.7ex\hbox{E}\kern-.125emX}}
\begin{document}
\newcommand{\RomanNumeralCaps}[1]
    {\MakeUppercase{\romannumeral #1}}
\def\BigRoman{\uppercase\expandafter{\romannumeral\number\count 255 }}
\def\Romannumeral{\afterassignment\BigRoman\count255=}

\title{On the Performance of Deep Learning-based Data-aided Active User Detection for GF-SCMA System}

\author{\IEEEauthorblockN{Minsig Han\IEEEauthorrefmark{1},
Ameha T. Abebe\IEEEauthorrefmark{2}, and
Chung G. Kang\IEEEauthorrefmark{1}}
\IEEEauthorblockA{\IEEEauthorrefmark{1}School of Electrical Engineering, Korea University, Seoul, Republic of Korea\\
\IEEEauthorrefmark{2}Samsung Research, Seoul, Republic of Korea\\
Email: \IEEEauthorrefmark{1}\{als4585, ccgkang\}@korea.ac.kr,
\IEEEauthorrefmark{2}amehat.abebe@samsung.com}}
\maketitle


\begin{abstract}
The recent works on a deep learning (DL)-based joint design of preamble set for the transmitters and data-aided active user detection (AUD) in the receiver has demonstrated a significant performance improvement for grant-free sparse code multiple access (GF-SCMA) system. The autoencoder for the joint design can be trained only in a given environment, but in an actual situation where the operating environment is constantly changing, it is difficult to optimize the preamble set for every possible environment. Therefore, a conventional, yet general approach may implement the data-aided AUD while relying on the preamble set that is designed independently rather than the joint design. In this paper, the activity detection error rate (ADER) performance of the data-aided AUD subject to the two preamble designs, i.e., independently designed preamble and jointly designed preamble, were directly compared. Fortunately, it was found that the performance loss in the data-aided AUD induced by the independent preamble design is limited to only 1dB. Furthermore, such performance characteristics of jointly designed preamble set is interpreted through average cross-correlation among the preambles associated with the same codebook (CB) (average intra-CB cross-correlation) and average cross-correlation among preambles associated with the different CBs (average inter-CB cross-correlation).
\end{abstract}

\begin{IEEEkeywords}
Grant-free access, deep learning, active user detection
\end{IEEEkeywords}

\section{Introduction}

In grant-free sparse code multiple access (GF-SCMA), the base station (BS) cannot determine which users are actively transmitting its data stream. Accordingly, GF users are preassigned a common preamble set and simultaneously transmit a randomly selected preamble from the preamble set along with the SCMA-based data stream. Then, active user detection (AUD) process becomes essential in the BS. Accordingly, a deep learning (DL)-based approaches have been recently employed for the joint design of preamble set for the transmitters and AUD in the receiver for GF-SCMA system [1]-[2].

Unlike the conventional preamble-based AUD, data-aided AUD has been considered for the autoencoder (AE) [1], which jointly optimizes a preamble set for the GF users and data-aided AUD in the BS, while demonstrating 3 to 5dB gain at the target activity detection error rate (ADER) of $10^{-3}$. In the joint optimization process, data-aided AUD is optimized assuming that a codebook (CB) uniquely associated with the preamble is exploited for the AUD process, and the preamble set is simultaneously optimized subject to the data-aided AUD. It is opposed to the conventional design that aims at minimizing the cross-correlation between any two preambles as they are determined independently regardless of the AUD process in the receiver side.

However, the jointly designed preamble set for DL-based data-aided AUD in [1] can be optimized only by training the AE in a given environment, which makes it challenging to be used in a practical situation where the operating environment is constantly varying. If the AE for joint design cannot be fully trained in practice, we have no choice but to consider the preamble designed independently at the transmitter side while giving up ADER performance. For the independent preamble set design under consideration, cross-correlation between the preambles tends to uniform, leading to a \textit{homogeneous} preamble set. However, the cross-correlation characteristic according to the joint design is significantly different, because the correlation between preambles depends on the detection characteristic associated with the CB, leading to a \textit{heterogenous} preamble set.

The primary purpose of this paper is to examine the ADER performance when an independently designed preamble set is applied to the data-aided AUD. Fortunately, it was found that the performance loss in the data-aided AUD induced by the independent preamble design is limited to only 1dB, as compared to 3 to 5dB gain with the joint design. It implies that the AE can be trained only for the receiver side, hopefully making it more practical, e.g., through online training even in the steadily varying environment. Furthermore, we attempt to analyze a performance characteristic of jointly designed preamble set, which allows for understanding the performance difference between two different implementation approaches. In particular, its performance characteristic is interpreted through average cross-correlation among the preambles associated with the same codebook (CB) (average intra-CB cross-correlation) and average cross-correlation among preambles associated with the different CBs (average inter-CB cross-correlation)

The rest of this paper is organized as follows. In Section II, we present a system model for GF-SCMA. Section III briefly discuss the DL-based design for independently designed preamble set and jointly designed preamble set while comparing the ADER performance of the two designs with each other. Lastly in Section IV, we provide the analysis on cross-correlation characteristics of the jointly designed preamble set for GF-SCMA with data-aided AUD. Finally, conclusion is made in Section V.

\section{System Model for Grant-free SCMA System}
In the GF-SCMA system, a total number of $N$ MTC users indexed by $n\in \{0,1,\ldots ,N-1\}$ are contending at the same time-frequency resource. At each transmission opportunity, only some of the $N$ users are actively transmitting its data. The activity of user $n$ is indicated by an activity indicator ${{\delta }_{n}}$, which can be modelled by the Bernoulli random variable with activity probability ${{p}_{n}}$. The activity indicators for all users constitutes a random vector, represented as $\bm{\delta }={{[{{\delta }_{0}},{{\delta }_{1}},\cdots ,{{\delta }_{N-1}}]}^{T}}$. Active user symbol of ${{\log }_{2}}M$ bits is encoded to a ${{K}_{d}}$-dimensional complex CB of size $M$. The ${{K}_{d}}$-dimensional complex codewords of the CB are sparse vectors with ${{N}_{m}}$ non-zero elements (${{N}_{m}}<{{K}_{d}}$)

A mapping matrix, that can be represented by a factor graph, maps the ${{N}_{m}}$ non-zero dimensions to the ${{K}_{d}}$-dimensional complex domain [3]-[4]. The SCMA encoder contains $J$ separate layers, which correspond to symbol nodes in the factor graph. Each layer is associated with a constellation set with $M$ alphabets of length ${{K}_{d}}$. The mapping matrix maps the ${{N}_{m}}$-dimensional constellation points to SCMA codewords to form the codeword set, which are multiplexed over ${{K}_{d}}$ shared orthogonal resources, e.g., OFDMA subcarriers. As $J$ layers are overloaded over ${{K}_{d}}$ resources, the overloading factor is given as $J/{{K}_{d}}$. A constellation set associated with the $j$-th layer of the SCMA encoder forms a CB, denoted as ${{\mathcal{C}}_{j}}$.

Let $\bar{\mathcal{C}}$ denote a set of $J$ CBs for the GF-SCMA system, i.e., $\bar{\mathcal{C}}=\{{{\mathcal{C}}_{0}},{{\mathcal{C}}_{1}},\ldots ,{{\mathcal{C}}_{J-1}}\}$, from which a CB is randomly selected for each of the active users. Note that the number of CBs is much smaller than the number of users, i.e., $J\ll N$. Therefore, the CB will be reused among users. However, even if they have employed the same CB, their transmissions may still be identified with their channel. [5] More specifically, when the two users select the different preambles and their channel is sufficiently different and reliably estimated, the multi-user detection (MUD) block in the BS is still able to detect the data streams, even if two users select the same CBs for their data transmission [6]. It is attributed to the fact that each CB selected by the active user can be uniquely identified by the individual channel that can be estimated through its own unique preamble. In other words, both CBs employed by users and their channel are jointly considered as the signatures for MUD. Thus, a tuple of SCMA CB and preamble effectively increases the number of contention resources, which allows for reducing the collision rate for GF-SCMA.

A contention transmission unit (CTU) is defined as a combination of time-frequency resources, SCMA CBs, and preamble sequences in the GF-SCMA system. When each CB is associated with $L$ preamble sequences, there are ${{N}_{R}}=J\times L$ CTUs in a given time-frequency region [5]. The GF-access users select one of these CTUs to transmit their data signals.

Notably, we assume that all preambles employ the round-robin CB association, i.e., CB with index $\nu (n)=\bmod (n,J)$, denoted as ${{\mathcal{C}}_{\nu (n)}}$, associated to a preamble with index $n$ [6]. Let ${{\mathbf{p}}^{(n)}}\in {{\mathbb{C}}^{{{K}^{(p)}}}}$ represent a $n$-th preamble sequence where ${{K}_{p}}$ denotes its dimension. In the current studies, as we aim at the preamble design for data-aided AUD, it is sufﬁcient to consider ${{N}_{R}}$ active users only, each assigned to a unique preamble, i.e., $N={{N}_{R}}$. In other words, our design considers a situation in which a maximum of $N={{N}_{R}}$ users are active without collision [1]. Therefore, $N$ is interchangeably used as a total number of users and a total number of preambles. Furthermore, in Fig. 1, the $n$-th user is uniquely assigned to preamble ${{\mathbf{p}}^{(n)}}$ and its associated CB ${{\mathcal{C}}_{\nu (n)}}$ for its CTU transmission.

\begin{figure}[t]
\centering
\includegraphics[width=0.8\linewidth,height=2.9in]{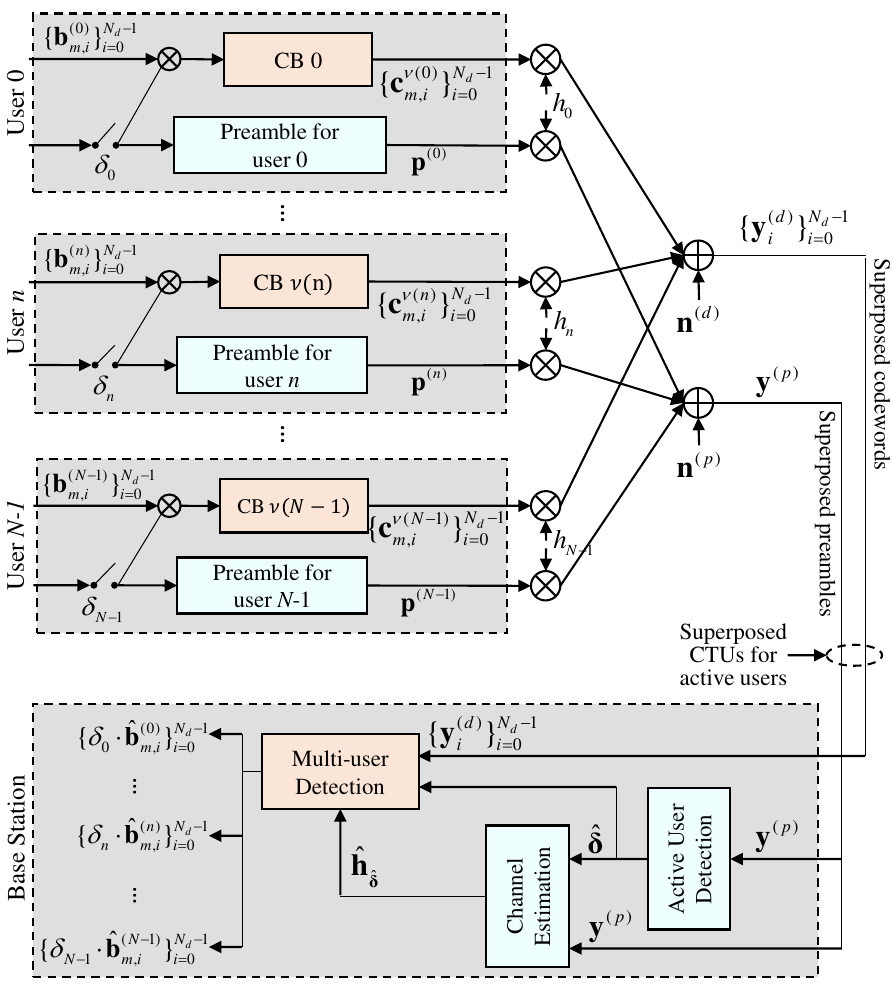}
\vspace*{-5pt}
\captionsetup{justification=raggedright, singlelinecheck=false}
\caption{Transceiver model for GF-SCMA system}
\label{fig_sim}
\vspace*{-10pt}
\end{figure}

Meanwhile, a data bit sequence of the active user is divided into ${{N}_{d}}$ blocks, each with ${{\log }_{2}}M$ bits. Then, each block is encoded with its own pre-determined CB. Let $\mathbf{b}_{i}^{(n)}\in {{\mathbb{B}}^{{{\log }_{2}}M}}$ denote the $i$-th block of user $n$, which is encoded with a codeword (CW), denoted as $\mathbf{c}_{i}^{\nu (n)}\in {{\mathcal{C}}_{\nu (n)}}$. In this case, CTU for user $n$ can be represented by $\left[ {{\mathbf{p}}^{(n)}}\left| \mathbf{c}_{1}^{\nu (n)},\mathbf{c}_{2}^{\nu (n)}, \right.\cdots ,\mathbf{c}_{{{N}_{d}}}^{\nu (n)} \right]$. Assuming that all CTUs are experiencing a flat fading channel, a CTU for user $n$ is subject to a single fading channel coefficient ${{h}_{n}}$ which constitutes a channel vector, represented as $\mathbf{h}={{[{{h}_{0}},{{h}_{1}},\ldots ,{{h}_{N-1}}]}^{T}}$. Consequently, the received signals, ${{\mathbf{y}}^{(p)}}$ and $\mathbf{y}_{i}^{(d)}$, for the preamble and the $i$-th CW superposed over the given resources, are expressed respectively as
\begin{equation}
{{\mathbf{y}}^{(p)}}=\sum\nolimits_{n=0}^{N-1}{{{\delta }_{n}}{{h}_{n}}{{\mathbf{p}}^{(n)}}}+{{\mathbf{n}}^{(p)}},
\label{eq}
\end{equation}
and
\begin{equation}
\mathbf{y}_{i}^{(d)}=\sum\nolimits_{n=0}^{N-1}{{{\delta }_{n}}{{h}_{n}}\mathbf{w}_{i}^{\nu (n)}}+\mathbf{n}_{i}^{(d)},\text{  }i\text{=}1,\ldots ,{{N}_{d}}.
\label{eq}
\end{equation}

Fig. 1 presents a transceiver model under consideration [1], in which AUD is performed only using (1) to detect an activity vector, represented as $\bm{\hat{\delta }}={{[{{\hat{\delta }}_{0}},{{\hat{\delta }}_{1}},\ldots ,{{\hat{\delta }}_{N-1}}]}^{T}}$. Then, channel coefficients are estimated for the active users by a CE process. The CE process yields an estimated channel coefficient of active user $n$, denoted as ${{\hat{h}}_{n}}$. Then, an estimated channel vector for active users is expressed as ${{\mathbf{\hat{h}}}_{{\mathbf{\hat{\delta }}}}}={{[{{\delta }_{0}}{{\hat{h}}_{0}},{{\delta }_{1}}{{\hat{h}}_{1}},\ldots ,{{\delta }_{N-1}}{{\hat{h}}_{N-1}}]}^{T}}$. Finally, active users’ transmitted bit sequences are detected by MUD in the BS.

\begin{figure}[t]
\centering
\vspace*{5pt}
\includegraphics[width=\linewidth]{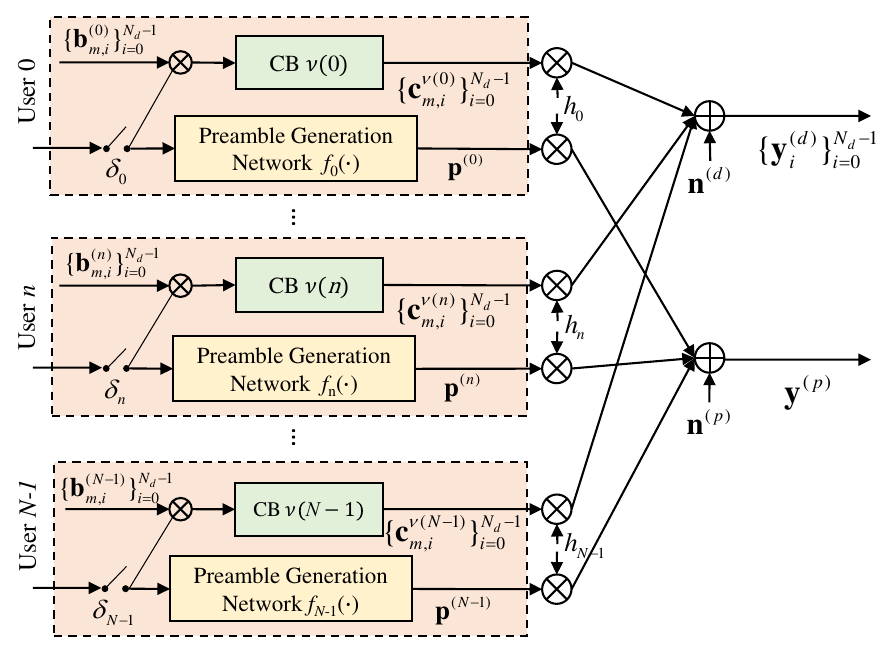}
\vspace*{-15pt}
\captionsetup{justification=raggedright, singlelinecheck=false}
\caption{A common encoder structure with the PGN}
\label{fig_sim}
\vspace*{-5pt}
\end{figure}

\section{Joint Design of Preamble Set: Heterogeneous vs. Homogeneous Preamble Set}
\subsection{AE implementation for the preamble set design}

In [1], a DL-based end-to-end design is proposed to jointly optimize a set of the $N$ preamble sequences and AUD at the receiver at the same time such that ADER can be minimized. It employs the AE for the end-to-end design, which is comprised of preamble generation networks (PGN), an active user detection network (AUDN), and a user activity extraction network (UAEN).  Note that the PGN in Fig. 2(a) can be expressed as
\begin{equation}
{{f}_{n}}({{\delta }_{n}};\bm{\theta }_{n}^{(f)})=\left\{ \begin{matrix}
   {{\mathbf{p}}^{(n)}},\text{  }{{\delta }_{n}}=1,  \\
   \mathbf{0},\text{   }{{\delta }_{n}}=0,  \\
\end{matrix} \right.
\label{eq}
\end{equation}
where $\bm{\theta }_{n}^{(f)}$ represent the weight and bias of the $n$-th PGN. Note that preambles are associated with one of the $J$ SCMA CBs.

Fig. 2 shows a common encoder structure with the PGN under consideration, in which the preamble sequence will be jointly designed with the either preamble-based AUD in Fig. 3(a) or data-aided AUD in Fig. 3(b) in the decoder. Meanwhile, Fig. 3 presents the decoder structures with AUD networks for the end-to-end preamble set design. For the conventional DL-based joint designs [2], AUD replies on the received preamble signal only as illustrated in Fig. 3(a) [7]-[9]. The preamble-based AUDN structure therein takes the superposed preamble ${{\mathbf{y}}^{(p)}}$ only as its input. Denoting its weight and bias by ${{\bm{\theta }}^{({{g}_{1}})}}$, it is represented as 
\begin{equation}
{{g}_{1}}({{\mathbf{y}}^{(p)}};{{\bm{\theta }}^{({{g}_{1}})}})=\bm{\hat{\delta }}
\label{eq}
\end{equation}
When all user’s activity probabilities are the same, i.e., ${{p}_{i}}={{p}_{j}},\forall j\ne i$, the cross-correlations of the preambles are optimized to be equally minimized. This is because PGN does not have any preconditions for distinguishing specific preambles so as to have the different cross-correlations. In other words, PGNs in the AE tries homogenously minimize cross-correlation of the $N$ preambles in a preamble set.

\begin{figure}[t]
\centering
\begin{subfigure}[b]{\linewidth}
\centering
\vspace*{5pt}
\includegraphics[width=\linewidth,height=1.2in]{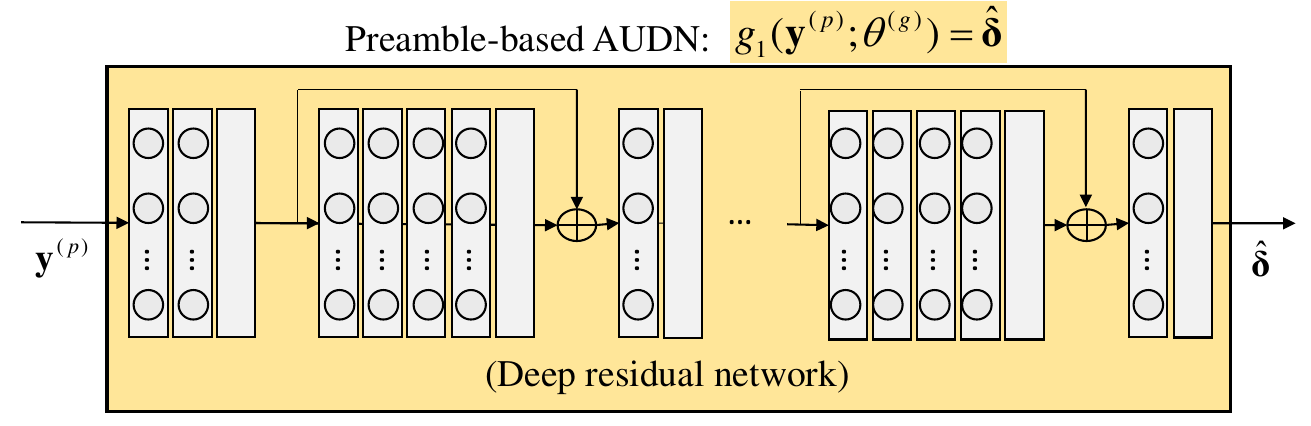}
\label{fig_sim}
\vspace*{-15pt}
  \caption{Preamble-based AUD network}
\end{subfigure}
\begin{subfigure}[b]{\linewidth}
\centering
\includegraphics[width=0.99\linewidth,height=1.2in]{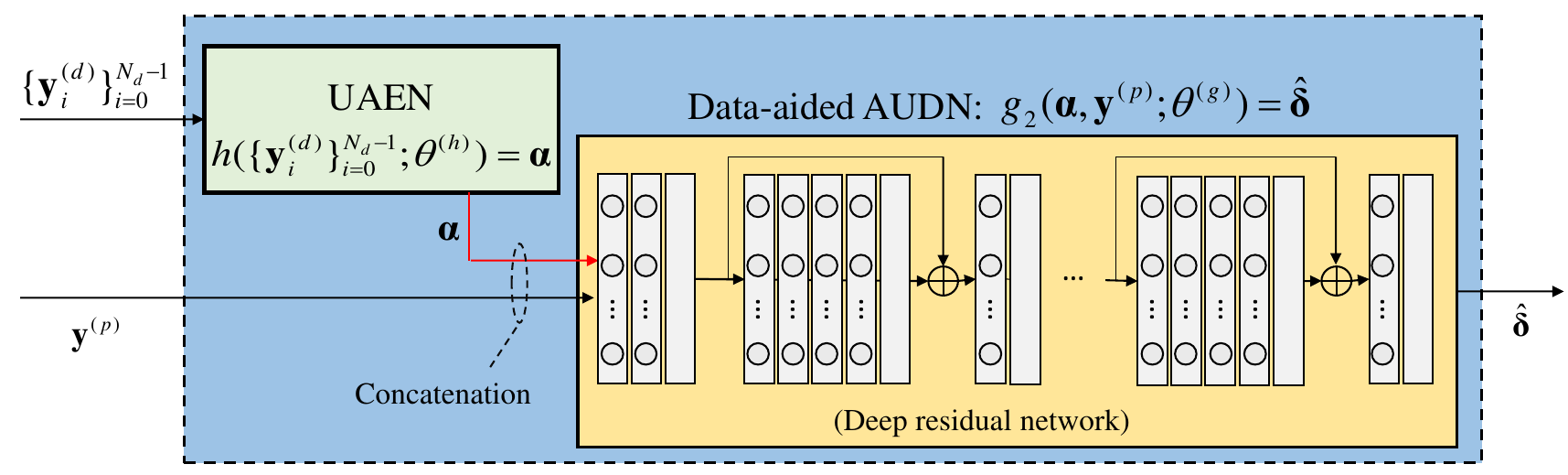}
\label{fig_sim}
\vspace*{-15pt}
  \caption{Data-aided AUD network} 
\end{subfigure}
\vspace*{-15pt}
\caption{Decoder structure with AUDN for the AE-based preamble design}
\vspace*{-10pt}
\end{figure}

Meanwhile, Fig. 3(b) presents the decoder structure with data-aided AUDN, which exploits both superposed preamble signals and the superposed data stream to detect the active users. As proposed in [1], this particular structure employs a user activity extraction network (UAEN), which extracts a-priori users’ activity information vector $\mathbf{\alpha }$ for the AUDN. It is important to note that the UAEN may not detect the active users only from $\{\mathbf{y}_{\ell }^{(d)}\}_{l=0}^{{{N}_{d}}-1}$ simply because the SCMA CBs are reused among the users. However, UAEN extracts a-priori activity information to indirectly know which CB is active in the current transmission. The data-aided AUD under consideration is to perform AUD exploiting the extracted information and the received preamble signal at the same time. Denoting the weight and bias of the UAEN by ${{\bm{\theta }}^{(h)}}$ and a-priori activity information for the data-aided AUDN by $\mathbf{\alpha }$, the UAEN in Fig. 3(b) is represented as 
\begin{equation}
h(\{\mathbf{y}_{i}^{(d)}\}_{i=0}^{{{N}_{d}}-1};{{\bm{\theta }}^{(h)}})=\bm{\alpha }
\label{eq}
\end{equation}
Furthermore, the data-aided AUDN that takes both a-priori activity information and superposed preamble can be expressed as follows: 
\begin{equation}
{{g}_{2}}(\bm{\alpha },{{\mathbf{y}}^{(p)}};{{\bm{\theta }}^{({{g}_{2}})}})=\bm{\hat{\delta }}
\label{eq}
\end{equation}
where ${{\bm{\theta }}^{({{g}_{2}})}}$ is the weight and bias of the data-aided AUDN. We note that the cross-correlation properties for the preamble sets that are jointly designed with the data-aided AUDN are not known clearly at the moment. However, as they depend mainly on the a-priori activity information, the cross-correlation may vary between preambles associated with the same CB and those associated with the different CBs, i.e., depending on whether the same CBs are selected or not.

\subsection{ADER performance analysis}
In this subsection, we present the simulation results to compare the ADER performance of jointly design preamble set and independent preamble set with each other. For simulation, the SCMA CB with $J=6$, ${{K}^{(d)}}=4$, ${{N}^{(d)}}=2$, and $M=4$ is assumed [3]. The preamble set is assumed to have $N$= 48, 60, or 72 preambles, i.e., the number of preambles associated with each of $J=6$ SCMA CB are $L$= 8, 10, and 12, respectively. The number of CWs exploited by UAEN is ${{N}_{d}}=16$ and the length of preamble sequence is given as ${{K}^{(p)}}=16$. We consider homogenous activity where every preamble has the same activity probabilities. In this case, let ${{N}_{a}}$ denote the expected number of active users. To fairly compare the ADER of different preamble set, we set ${{N}_{a}}=3$, by setting the homogenous activity probabilities as 0.0625, 0.05, and 0.04166 for the case of $N$= 48, 60, or 72 respectively. The detailed system parameters and hyperparameters for the AE in Fig. 2 and Fig. 3 follows those in [1].

Fig. 4 presents the ADER performance for the different AUD schemes as varying the number of users and a preamble size. We are interested in the performance gain of data-aided AUD with jointly design preamble and independent preamble over the preamble-based AUD. As already observed in [1], a joint design for the DL-based data-aided AUD scheme achieves a combined gain of 3 to 5dB gain over the preamble-based AUD. Meanwhile, its gain with independent preamble has been reduced to 2 to 4dB, depending on the range of SNR and the number of preambles $N$. Fortunately, it corresponds to the performance loss of only 1dB or less compared to independently designed preamble in ADER of ${{10}^{-3}}$ in high SNR region. It implies that our proposed DL-based data-aided AUD can be still useful without resorting to the joint design of preambles with the AUD. In fact, it makes the DL-based design approach more practical, because the AE can be trained online over the air, while avoiding the offline supervised training, under the steadily varying environment.

In the following section, we attempt to analyze a performance characteristic of jointly designed preamble set, hopefully allowing for understanding the characteristics that lead to performance differences between two different setups.

\begin{figure}[t]
\centering
\begin{subfigure}[b]{\linewidth}
\centering
\includegraphics[width=0.8\linewidth]{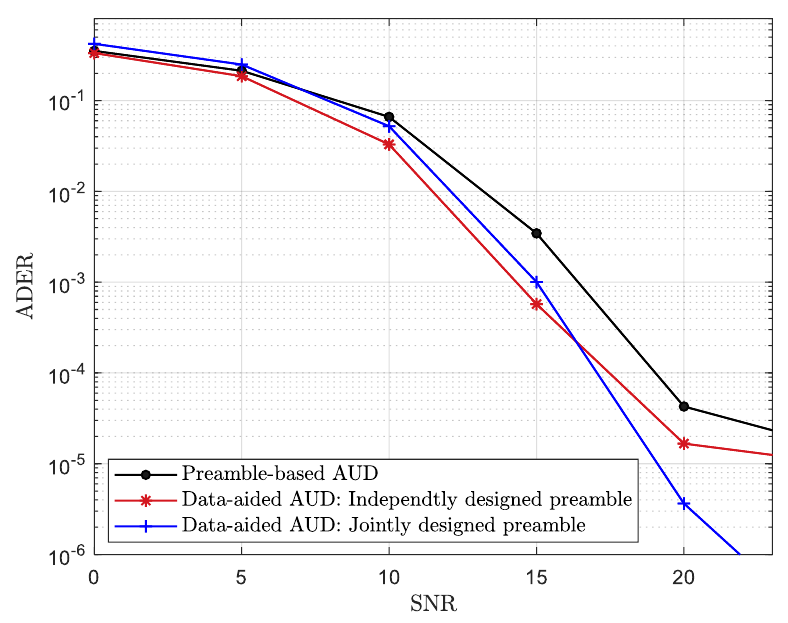}
\label{fig_sim}
\vspace*{-5pt}
  \caption{$N$=48 \& $L$=8} 
\end{subfigure}
\begin{subfigure}[b]{\linewidth}
\centering
\includegraphics[width=0.8\linewidth]{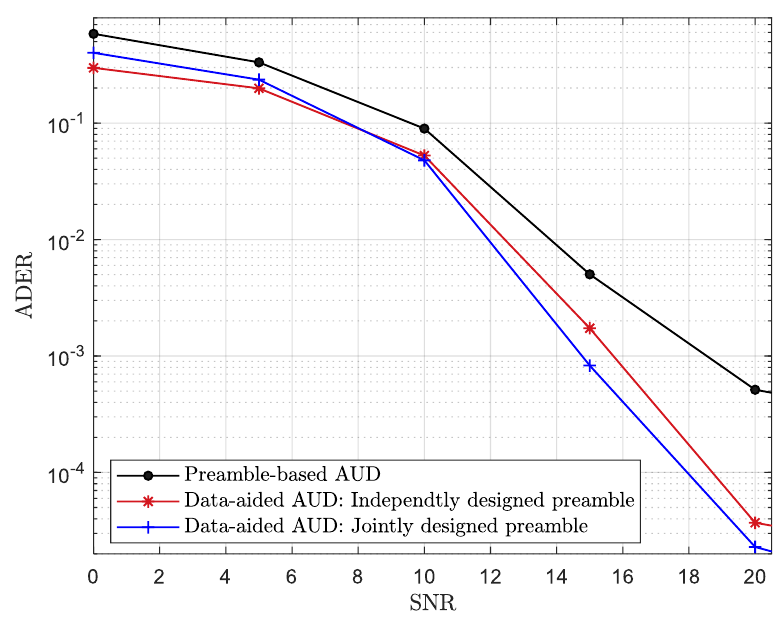}
\label{fig_sim}
\vspace*{-5pt}
  \caption{$N$=60 \& $L$=10} 
\end{subfigure}
\begin{subfigure}[b]{\linewidth}
\centering
\includegraphics[width=0.8\linewidth]{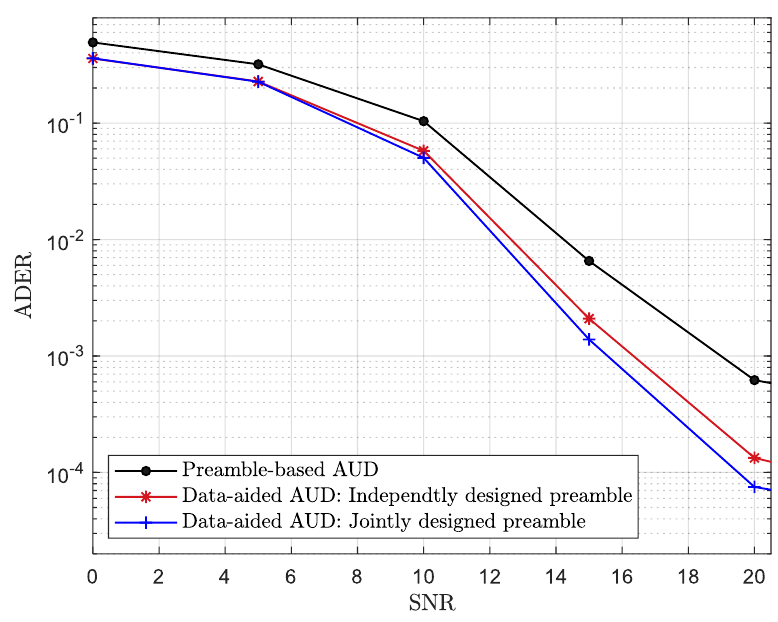}
\label{fig_sim}
\captionsetup{font=small,justification=centering}
\vspace*{-5pt}
  \caption{$N$=72 \& $L$=12} 
\end{subfigure}
\vspace*{-15pt}
\captionsetup{justification=raggedright, singlelinecheck=false}
\caption{ADER performance for the different AUD schemes}
\vspace*{-15pt}
\end{figure}

\begin{figure*}[t]
\begin{subfigure}[b]{0.325\textwidth}
\includegraphics[width=0.97\linewidth]{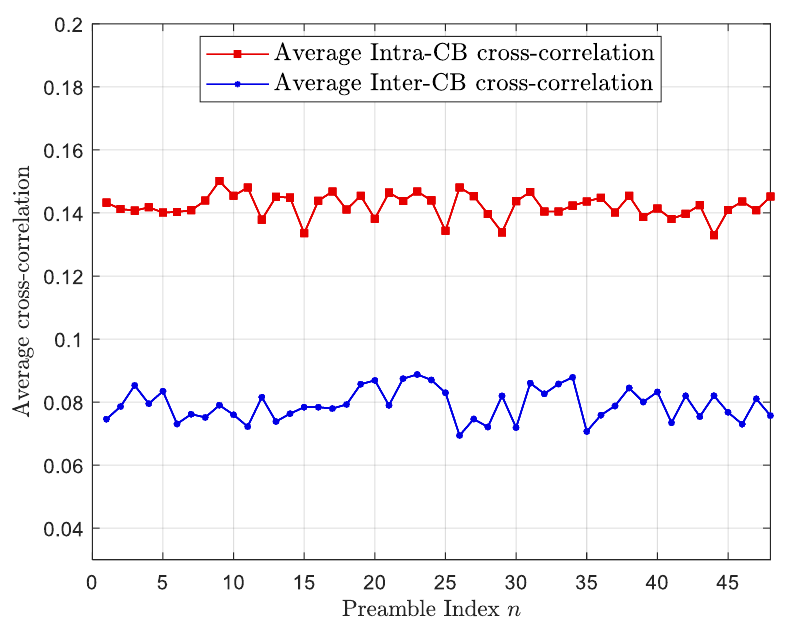}
\label{fig_sim}
  \caption{$N$=48 \& $L$=8} 
\end{subfigure}
\begin{subfigure}[b]{0.325\textwidth}
\includegraphics[width=0.97\linewidth]{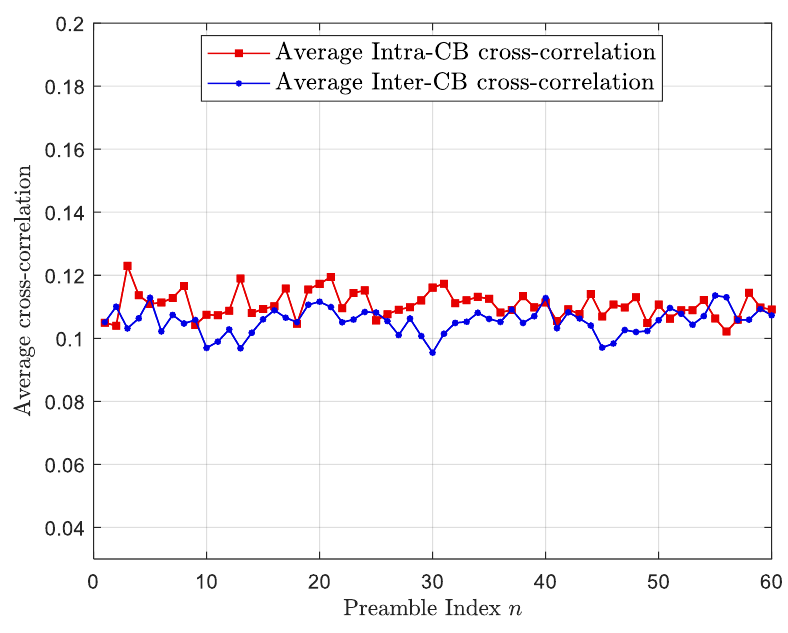}
\label{fig_sim}
  \caption{$N$=60 \& $L$=10} 
\end{subfigure}
\begin{subfigure}[b]{0.325\textwidth}
\includegraphics[width=0.97\linewidth]{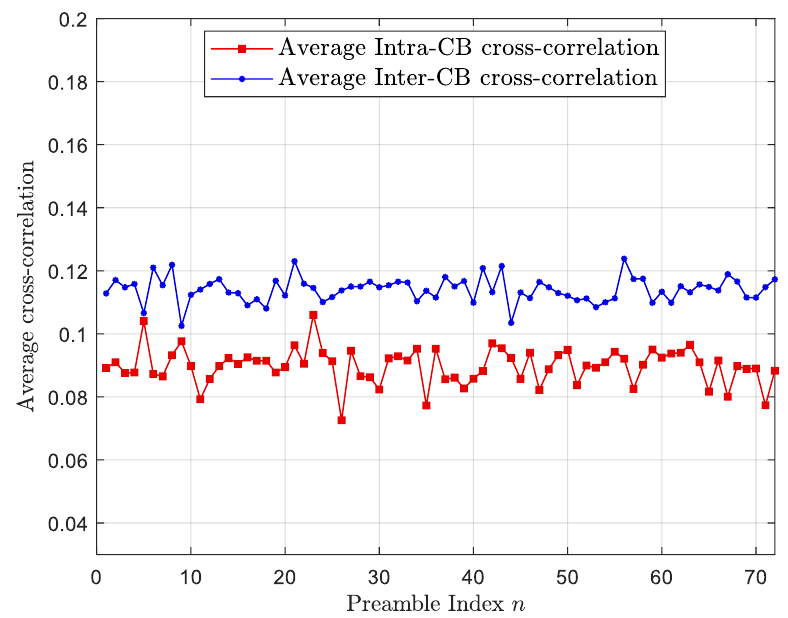}
\label{fig_sim}
  \caption{$N$=72 \& $L$=12} 
\end{subfigure}
\vspace*{-5pt}
\captionsetup{justification=raggedright, singlelinecheck=false}
\caption{Average intra-CB and inter-CB cross-correlation as varying the number of users and preamble size}
\vspace*{-10pt}
\end{figure*}

\section{On the Performance Characteristics of Heterogeneous Preamble Set}
In order to understand the results in the previous section, the performance characteristics must be investigated for the preamble sequence. As discussed in [2], it is desirable that the cross-correlation between any two sequences is as small as possible to improve the preamble-based AUD process. Let $\mathbf{p}_{n}^{{}}$ a ${{K}_{p}}$-dimensional vector to represent the $n$-th preamble sequence in the given preamble set, i.e., $\mathbf{p}_{n}^{T}=[{{p}_{n,1}},{{p}_{n,2}},\cdots ,{{p}_{n,{{K}_{p}}}}]$. Then, its cross-correlation between the $n$-th and $m$-th preamble sequences in the set is given as ${{R}_{n,m}}\triangleq <{{\mathbf{p}}_{n}},{{\mathbf{p}}_{m}}>=\sum\limits_{i=1}^{{{K}_{p}}}{{{p}_{n,i}}{{p}_{m,i}}}$.

The AUD performance of the given preamble set can be inferred by the average cross-correlation between all the other preambles in the given preamble set. Let $\mathbf{p}_{l}^{(j)}$ denote the $l$-th preamble associated with the $j$-th CB. The average cross-correlation of preamble $\mathbf{p}_{l}^{(j)}$, denoted as $\bar{R}_{\ell }^{(j)}$, is given as
\[\bar{R}_{\ell }^{(j)}=\frac{1}{N-1}\left\{ \sum\limits_{k=0}^{J-1}{\sum\limits_{m=0}^{L-1}{\left| \left\langle \mathbf{p}_{\ell }^{(j)},\mathbf{p}_{m}^{(k)} \right\rangle  \right|}}-\left| \left\langle \mathbf{p}_{\ell }^{(j)},\mathbf{p}_{\ell }^{(j)} \right\rangle  \right| \right\},\]
\begin{equation}
\ell =0,1,\cdots ,L-1; j=0,1,\cdots ,J-1 
\end{equation}

As (7) averages out the cross-correlation among the entire preambles except itself, it can be a useful metric to measure the performance characteristics for a homogeneous preamble set. However, it does not capture the cross-correlation for the heterogeneous preamble sets designed for data-aided AUD [1], which varies by whether the preambles belong to the same CB or not. In order to reveal their performance characteristics, we need to consider their cross-correlation among the preambles in the same CB and those in the different CBs. Toward this end, we consider two different average cross-correlations among the preamble sequences, one for intra-CB cross-correlation denoted as $\bar{R}_{j,l}^{\operatorname{intra}}$ and the other for inter-CB cross-correlation denoted as $\bar{R}_{j,l}^{\operatorname{inter}}$. They are defined respectively as
\begin{equation}
\bar{R}_{j,l}^{\operatorname{intra}}=\frac{1}{L-1}\sum\limits_{m=0,m\ne l}^{L-1}{\left| \left\langle \mathbf{p}_{l}^{(j)},\mathbf{p}_{m}^{(j)} \right\rangle  \right|}
\end{equation}
and
\begin{equation}
\resizebox{\columnwidth}{!} {
$\bar{R}_{j,l}^{\operatorname{inter}}=\frac{1}{N-L}\left\{ \sum\limits_{k=0}^{J-1}{\sum\limits_{m=0}^{L-1}{\left| \left\langle \mathbf{p}_{\ell }^{(j)},\mathbf{p}_{m}^{(k)} \right\rangle  \right|}}-\sum\limits_{m=0}^{L-1}{\left| \left\langle \mathbf{p}_{l}^{(j)},\mathbf{p}_{m}^{(j)} \right\rangle  \right|} \right\}.$
}
\end{equation}

Fig. 5 present the average intra-CB and inter-CB cross-correlation, $\bar{R}_{j,l}^{\operatorname{intra}}$ and $\bar{R}_{j,l}^{\operatorname{inter}}$, as varying the number of preamble size. An index $n$ therein represents one for the $l$-th preamble associated with the $j$-th CBs, which is equivalently given as $n=L\cdot j+l$. First of all, it is observed in Fig. 5 that all values of cross-correlation between the different CBs (differentiated by the different indices) are similar to each other. This observation validates the fact that the preamble-CB association leads to the heterogenous preamble sets for our joint design. Furthermore, as the total number of users $N$ increase from 48 to 72, the average intra-CB cross-correlation is decreasing, and average inter-CB cross-correlation are increasing.

\begin{figure}[t]
\centering
\includegraphics[width=0.8\linewidth]{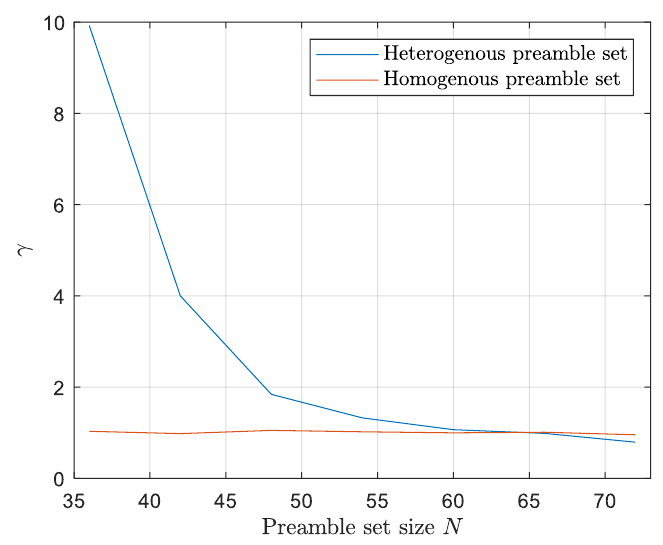}
\captionsetup{justification=raggedright, singlelinecheck=false}
\caption{Heterogeneity as varying the preamble size: $J$ = 6}
\vspace*{-15pt}
\label{fig_sim}
\end{figure}

Meanwhile, we note that the homogeneity of preamble sets is characterized by the distribution of $\bar{R}_{j,l}^{\operatorname{intra}}$ and $\bar{R}_{j,l}^{\operatorname{inter}}$, i.e., they tend to be equal to each other for the homogeneous preambles. Toward that end, we consider the average value of $\bar{R}_{j,l}^{\operatorname{intra}}$ and $\bar{R}_{j,l}^{\operatorname{inter}}$ across the entire preamble sets, i.e., defined as ${{\bar{R}}_{\text{intra}}}=\sum\limits_{j=0}^{J-1}{\sum\limits_{l=0}^{L-1}{\bar{R}_{j,l}^{\operatorname{intra}}}}$ and ${{\bar{R}}_{\text{inter}}}=\sum\limits_{i=0}^{J-1}{\sum\limits_{l=0}^{L-1}{\bar{R}_{j,l}^{\operatorname{inter}}}}$. Then, ${{\bar{R}}_{\text{intra}}}$ and ${{\bar{R}}_{\text{inter}}}$ can be compared by a ratio of $\gamma ={{\bar{R}}_{\text{intra}}}/{{\bar{R}}_{\text{inter}}}$ as a metric of measuring the homogeneity of the preamble sets obtained by two different designs. More specifically, the preamble sets tend to be homogeneous as $\gamma$ approaches one.

As shown in Fig. 6, as the total number of users increases, the average intra-CB cross-correlation decreases while increasing the average inter-CB cross-correlation. It can be more clearly observed by presenting the heterogeneity in terms of the ratio $\gamma ={{\bar{R}}_{\text{intra}}}/{{\bar{R}}_{\text{inter}}}$, as shown with $J$ = 6 in Fig. 6. We observe that the heterogeneity of the preamble sets decreases as increasing the number of preambles. In particular, as the number preamble associated with each CB becomes extremely small, intra-CB cross correlation becomes large. For example, suppose only two preambles are associated with each of the $J$=6 CBs, i.e., $L$=2, and $N=J\cdot L=12$. Then, detection of these two preambles in the same CB leads to more likely association between the CB and preambles, i.e., increasing the average intra-CB cross-correlation more significantly. In other words, a-prior activity information obtained by CB activation becomes more significant as the number of preambles associated with each CB is reduced.

\section{Conclusion}
In this paper, the performance of the joint design for preamble set and data-aided AUD is explored. To confirm the performance gain associated with the preamble design, the performance of DL-based data-aided AUD is investigated with independently designed preamble set and jointly designed preamble sets. In conclusion, the independently designed preamble set suffers from performance degradation of only 1dB compared to the jointly designed heterogenous preamble for data-aided AUD. It implies that the AE can be trained only for the receiver side, hopefully making it more practical, at the loss of 1dB or less. For example, online over-the-air AE training can be employed under the steadily varying environment, rather than supervised learning for all possible predictable environments.

In the future work, it is worthwhile to investigate an online training scheme that can design instantaneous heterogenous preamble sets to be jointly optimized with data-aided AUD. Meanwhile, since this study consider only one SCMA CB set, multiple CB sets can be adopted to support the diversity over data-aided AUD process [6]. They enable to infer preamble activity more from CB activity detection as the number of associated preambles per CB is reduced.

\section*{Acknowledgment}
This work was supported in part by Institute of Information \& communications Technology Planning \& Evaluation (IITP) grant funded by the Korea government (MSIT) (No.2021-0-00467, Intelligent 6G Wireless Access System) and in part by the National Research Foundation of Korea (NRF) grant funded by the Korea government (MSIT) (No.2020R1A2C100998413).


\begin{thebibliography}{1}
\bibitem{IEEEhowto:kopka} 
M. Han, A. T. Abebe and C. G. Kang, "Data-aided Active User Detection with a User Activity Extraction Network for Grant-free SCMA Systems," arXiv Preprint, May. 2022, arXiv:2205.10780. \hskip 1em plus
  0.5em minus 0.4em\relax

\bibitem{IEEEhowto:kopka} 
N. Kim, D. Kim, B. Shim and K. B. Lee, "Deep Learning-Based Spreading Sequence Design and Active User Detection for Massive Machine-Type Communications," \emph{IEEE Wireless Commun. Lett.}, vol. 10, no. 8, pp. 1618-1622, Aug. 2021. \hskip 1em plus
  0.5em minus 0.4em\relax


\bibitem{IEEEhowto:kopka} 
M. Taherzadeh, H. Nikopour, A. Bayesteh and H. Baligh, "SCMA Codebook Design," in \emph{Proc. 2014 IEEE 80th Vehicular Technology Conference (VTC2014-Fall)}, vol. 3, no. 4, Dec. 2017, pp. 563-575. \hskip 1em plus
  0.5em minus 0.4em\relax

\bibitem{IEEEhowto:kopka} 
M. Han, H. Seo, A. T. Abebe and C. G. Kang, "Deep Learning-based Codebook Design for Code-Domain Non-Orthogonal Multiple Access Approaching Single-User Bit-Error Rate Performance," \emph{IEEE Trans. Cognitive Commun. Netw.}, vol. 8, no. 2, pp. 1159-1173, June 2022. \hskip 1em plus
  0.5em minus 0.4em\relax

\bibitem{IEEEhowto:kopka} 
K. Au et al., "Uplink contention based SCMA for 5G radio access," in \emph{Proc. 2014 IEEE Globecom Workshops (GC Wkshps)}, 2014, pp. 900-905. \hskip 1em plus
  0.5em minus 0.4em\relax

\bibitem{IEEEhowto:kopka} 
A. T. Abebe and C. G. Kang, "Grant-Free Uplink Transmission with Multi-Codebook-Based Sparse Code Multiple Access (MC-SCMA)," \emph{IEEE Access}, vol. 7, pp. 169853-169864, 2019. \hskip 1em plus
  0.5em minus 0.4em\relax

\bibitem{IEEEhowto:kopka} 
W. Kim, Y. Ahn and B. Shim, "Deep Neural Network-Based Active User Detection for Grant-Free NOMA Systems," \emph{IEEE Trans. Commun.}, vol. 68, no. 4, pp. 2143-2155, April 2020.  \hskip 1em plus
  0.5em minus 0.4em\relax

\bibitem{IEEEhowto:kopka} 
Y. Ahn, W. Kim and B. Shim, "Active User Detection and Channel Estimation for Massive Machine-Type Communication: Deep Learning Approach,"  \emph{IEEE Internet Things J.}, 2021, pp. 1-1. \hskip 1em plus
  0.5em minus 0.4em\relax

\bibitem{IEEEhowto:kopka} 
T. Sivalingam, S. Ali, N. H. Mahmood, N. Rajatheva and M. Latva-Aho, "Deep Learning-Based Active User Detection for Grant-free SCMA Systems," in \emph{Proc. 2021 IEEE 32nd Annual International Symposium on Personal, Indoor and Mobile Radio Communications (PIMRC)}, 2021, pp. 635-641. \hskip 1em plus
  0.5em minus 0.4em\relax






\end{thebibliography}
\end{document}